\documentclass[
aps,
pra,
superscriptaddress,
twocolumn,
a4paper,
floatfix,
longbibliography,
raggedbottom
]{revtex4-1}

\usepackage{amsmath}
\usepackage{amssymb}
\usepackage{bbm}
\usepackage{mathtools}
\usepackage{dsfont}
\usepackage{braket}

\usepackage{graphicx}
\usepackage[svgnames]{xcolor}

\usepackage{siunitx} 

\usepackage{ragged2e}
\usepackage{array}
\usepackage{tabularx}
\usepackage{booktabs}
\usepackage{textcomp}

\definecolor{cset-aps-blue}{RGB}{18,84,168}
\definecolor{cset-aps-limegreen}{RGB}{153,204,51}

\definecolor{cset-aps-blueberry}{RGB}{28,128,158}
\definecolor{cset-aps-turquoise}{RGB}{0,67,88}
\definecolor{cset-aps-limegreen}{RGB}{190,219,67}
\definecolor{cset-aps-darkblue}{RGB}{31,138,112}
\definecolor{cset-aps-yellow}{RGB}{255,225,25}
\definecolor{cset-aps-orange}{RGB}{253,116,0}
\definecolor{cset-aps-red}{RGB}{219,0,43}

\usepackage{tikz}

\usepackage{pgfplots}
\pgfplotsset{%
	every axis legend/.append style={%
		cells={anchor=west},
		at={(0.96,0.04)},
		anchor=south east,
		font=\scriptsize,
	},
	every axis/.append style={%
		yticklabel style={%
			/pgf/number format/fixed zerofill,
			/pgf/number format/precision=2},
	},
	width= \textwidth,
	height=8cm,
	xmajorgrids=false,
	xminorgrids=false,
	minor x tick num=1,
	colorbar style={yticklabel style={%
			/pgf/number format/precision=1,
			/pgf/number format/fixed},
		width=0.08*\pgfkeysvalueof{/pgfplots/parent axis width}}
}
\usepgfplotslibrary{external}	
\usepgfplotslibrary{colorbrewer}
\usepgfplotslibrary{colormaps}
\usepgfplotslibrary{groupplots}
\usepackage{xcolor}

\usetikzlibrary{decorations}

\newcommand{\D}[0]{\mathrm{d}}

\newcommand{\I}[0]{{\mathrm{i}}}
\newcommand{\e}[0]{{\mathrm{e}}}

\usepackage{hyperref}
\hypersetup{%
	colorlinks=true,
	linkcolor={cset-aps-red},
	linkbordercolor={cset-aps-red},
	filecolor={cset-aps-orange},
	filebordercolor={cset-aps-orange},
	citecolor={cset-aps-blue},
	citebordercolor={cset-aps-blue},
	urlcolor={cset-aps-blue},
	urlbordercolor={cset-aps-blue},
	menucolor={cset-aps-limegreen},
	menubordercolor={cset-aps-limegreen},
	breaklinks=true,
	pdfborderstyle={/S/U/W 2},
	pdfpagemode=UseOutlines,
	pdfstartpage={1},
}
\usepackage[capitalize]{cleveref}

\makeatletter
\def\maketitle{
	\@author@finish
	\title@column\titleblock@produce
	\suppressfloats[t]}
\makeatother

\newcommand{\affDLR}{\address{Institute of Quantum Technologies, German Aerospace Center (DLR), Wilhelm-Runge-Stra\ss e 10, D-89081 Ulm, Germany}}
\newcommand{\affHAN}{\address{Institut f{\"u}r Quantenoptik, Leibniz Universit{\"a}t Hannover,
		Welfengarten 1, D-30167 Hannover, Germany}}
\newcommand{\affULM}{\address{Institut f{\"u}r Quantenphysik and Center for Integrated Quantum
		Science and Technology (IQ\textsuperscript{ST}), Universit{\"a}t Ulm, Albert-Einstein-Allee 11, D-89069 Ulm, Germany}}

\begin{document}
	
	\title[Third-order double Raman diffraction]{Atomic Raman scattering: Third-order diffraction in a double geometry\\[1ex]
		\normalsize\normalfont{Published in \href{https://journals.aps.org/pra/abstract/10.1103/PhysRevA.102.063326}{Phys. Rev. A {\bfseries 102},  063326 (2020)}}}

	\author{Sabrina Hartmann} \thanks{sabrina.hartmann@uni-ulm.de}
	\affULM
	\author{Jens Jenewein}   
	\affULM
	\author{Sven Abend}   
	\affHAN
	\author{Albert Roura}   
	\affDLR
	\author{Enno Giese}   
	\affULM
	\affHAN
	
	\begin{abstract}
		In a retroreflective scheme with an atom initially at rest, atomic Raman diffraction adopts some of the properties of Bragg diffraction due to additional couplings to off-resonant momenta. As a consequence, double Raman diffraction has to be performed in a Bragg-type regime, where the pulse duration is sufficiently long to suppress diffraction into spurious orders. Taking advantage of this regime, double Raman allows for resonant higher-order diffraction. We study theoretically the case of third-order diffraction and compare it to first order as well as a sequence of first-order Raman pulses giving rise to the same momentum transfer as the third-order pulse.
		Moreover, we demonstrate that interferometry is possible and investigate amplitude and contrast of a third-order double-Raman Mach-Zehnder interferometer.
		In fact, third-order diffraction constitutes a competitive tool for the diffraction of ultracold atoms and interferometry based on large momentum transfer since it allows to reduce the complexity of the experiment as well as the total duration of the diffraction process compared to a sequence, at the cost of higher pulse intensities.
	\end{abstract}

	\maketitle
	\section{Introduction}
	Higher-order Bragg diffraction \cite{muller_atom_2008,giese2013double,kuber2016experimental,siem2020analytic,HU2017632} in combination with sequential pulses \cite{PhysRevLett.116.173601,chiow_$102ensuremathhbark$_2011} has become a standard tool for large-momentum-transfer (LMT) techniques to enhance the sensitivity of light-pulse atom interferometers \cite{kasevich_atomic_1991,kleinert_representation-free_2015}.
	However, with Raman diffraction \cite{PhysRevLett.66.2297,kasevich_atomic_1991,PhysRevA.45.342}, the other main mechanism, only sequential pulses \cite{PhysRevLett.103.080405,PhysRevLett.85.4498,PhysRevLett.114.063002} have routinely been employed so far. In this article, we extend Raman in a double-diffraction geometry \cite{PhysRevA.81.013617,PhysRevA.100.053618,PhysRevLett.115.013004} to also allow for higher-order diffraction, study the efficiency compared to a standard first-order sequence, and simulate a simple interferometer. Such a setup retains the possibility of state-selective detection, while being more efficient and less complex than a sequence of first-order pulses for narrow momentum distributions.
	
	Sequential pulses \cite{PhysRevLett.116.173601,chiow_$102ensuremathhbark$_2011, kovachy2015quantum, PhysRevLett.103.080405,PhysRevLett.85.4498,PhysRevLett.114.063002} and higher-order diffraction \cite{muller_atom_2008,giese2013double,kuber2016experimental,siem2020analytic,HU2017632} are some of the most common techniques used for LMT applications based on Bragg diffraction and are often combined with Bloch oscillations \cite{PhysRevLett.102.240402,PhysRevA.88.053620,abend_atom-chip_2016, pagel_bloch_2019,gebbe_twin-lattice_2019}.
	They are complemented by double diffraction \cite{giese2013double, PhysRevLett.116.173601, PhysRevA.81.013617,PhysRevA.100.053618,PhysRevLett.115.013004}, where an atom at rest diffracts in two opposite directions from two counterpropagating pairs of light fields, each involving two different frequencies. It is particularly well suited for experiments under microgravity conditions \cite{geiger2011detecting,muntinga_interferometry_2013,becker_space-borne_2018,aveline2020observation,frye_bose-einstein_2019,Aguilera_2014,Hogan2011} or for horizontal geometries \cite{canuel2019elgar,schubert2019scalable}, where a vanishing initial momentum arises naturally.
	Due to its symmetry, laser phases are not imprinted on the two branches of the interferometer, and similar noise sources are intrinsically suppressed \cite{PhysRevLett.103.080405,PhysRevLett.116.173601, giese2013double}. Even though many applications of double diffraction focus on Bragg, the geometry was first pioneered for Raman and is still used to date as one of the few LMT techniques for Raman diffraction, together with sequential pulses. However, one of the benefits of double Raman diffraction has not been explored so far, namely the possibility to scatter into higher diffraction orders. 
	
	In contrast to single Raman, which can be described as a closed two-level system, off-resonant couplings appear in single Bragg diffraction \cite{torii_mach-zehnder_2000,giese2015mechanisms}, limiting the operation to the Bragg regime with long pulse durations but at the same time allowing for higher-order diffraction for sufficiently long pulses \cite{muller2008atom}. The additional pair of light fields in double diffraction induces further off-resonant transitions for both Raman and Bragg diffraction.
	As a consequence, the application of Raman diffraction is restricted to a Bragg-type regime as well, where the pulses are so long that the effective Rabi frequency is much smaller than the frequency associated with the kinetic energy gained during the diffraction process \footnote{We use the expression \emph{Bragg-type regime} for Raman diffraction instead of \emph{Bragg regime}, to avoid any confusion that there exists also a Bragg regime for Raman diffraction.}. In double Bragg diffraction resonant and off-resonant couplings at the same momentum state appear causing a more complex diffraction behavior \cite{giese2013double,PhysRevA.101.053610}. However, these features do not arise in double Raman diffraction, which therefore constitutes a simpler diffraction mechanism.
	
	In this article we demonstrate that third-order double Raman diffraction with high efficiency and interferometry based on this mechanism are possible, although the process is more velocity selective than its first-order counterpart.
	However, for narrow momentum distributions like the ones associated with Bose-Einstein condensates (BECs) it can be a competitive alternative to a pulse sequence when the duration of the beam splitting process is limited, at the cost of higher pulse intensities.
	
	In Sec.~\ref{sec:FirstOrderDiffraction} we recall first-order double Raman diffraction with a Gaussian pulse shape as well as sequential Doppler-detuned single Raman diffraction with typical box-shaped pulses to calculate the efficiency of an LMT beam splitter.
	Such a combination of Gaussian and box-shaped pulses constitutes a good compromise between diffraction efficiency and overall duration of the sequence.
	We then perform in Sec.~\ref{sec:ThirdOrderDiffraction} an analysis of third-order double-Raman beam splitters and show that even though their efficiency is inherently worse than a comparable first-order pulse, it can be better than that of the sequence.
	In a similar manner, Sec.~\ref{sec.Mirror_processes} focuses on third-order double-Raman mirror processes, which turn out to be less efficient than the beam splitters. 
	Combining both mirror and beam splitters to a simple Mach-Zehnder interferometer, Sec.~\ref{sec.Interference_signal} demonstrates that third-order Raman diffraction is a suitable tool for atom interferometry by calculating the properties of interference signals and demonstrating phase coherence.
	We conclude with a brief discussion in Sec.~\ref{sec:Conclusion}.
	For completeness, the general set of differential equations for double Raman diffraction is given in Appendix~\ref{app:GeneralEquations} and energy shifts that appear in third-order diffraction with box-shaped pulses are detailed in Appendix~\ref{app.EnergyShifts}.

	\section{First-order diffraction}
	\label{sec:FirstOrderDiffraction}
	\subsection{Double Raman diffraction}
	\label{subsec:DoubleRaman}
	\begin{figure}[tb]
		\includegraphics[width = \columnwidth]{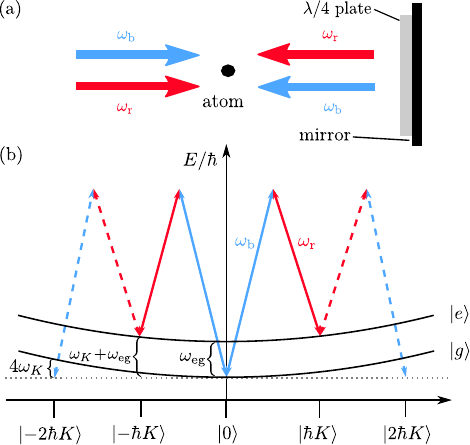}
		\caption{Schematic setup with an atom at rest ($p_0=0$) in a retroreflective geometry built by a $\lambda / 4$ plate and a mirror (a). The atom diffracts from two counterpropagating pairs of light fields, each consisting of two different frequencies, (upper and lower pair), absorbs a photon with frequency $\omega_\mathrm{b}$ and emits a photon with frequency $\omega_\mathrm{r}$ in opposite direction from each laser pair. This process causes a total recoil of $\pm \hbar K$ with $K =(\omega_\mathrm{b} + \omega_\mathrm{r})/c $ and by that leads to a gain of kinetic energy $\hbar \omega_K$. The energy-momentum diagram in (b) shows that such a process is resonant if the energy difference between the light fields $\hbar \Delta \omega \equiv \hbar (\omega_\mathrm{b} - \omega_\mathrm{r})$ equals the energy difference $\hbar \omega_\mathrm{eg}$ between the atomic ground $\ket{g}$ and excited state $\ket{e}$ in addition to the recoil energy $\hbar \omega_K$. Resonant processes start and end on the parabola (solid arrows), off-resonant processes are denoted by dashed arrows.}
		\label{fig:FirstOrderSetup}
	\end{figure}
	An atom at rest interacts with two strongly detuned pairs of light fields that move in opposite directions (with a detuning much larger than the linewidth), each one generated by two counterpropagating light fields of frequencies $\omega_\mathrm{b}$ and $\omega_\mathrm{r}$, see \cref{fig:FirstOrderSetup}(a). The pairs can be distinguished by their combination of polarization \cite{PhysRevLett.103.080405}, so that within a retroreflective setup, where both light fields are guided from one side to the atom and retroreflected at the other side, the polarizations have to be rotated by a $\lambda/4$ plate \cite{PhysRevLett.103.080405, PhysRevLett.116.173601} to suppresses spurious couplings. 
	The diffraction process can be understood in terms of absorbing a photon with frequency $\omega_\mathrm{b}$ and subsequently emitting a photon with frequency $\omega_\mathrm{r}$ in the opposite direction from each pair of light fields. This process causes a total momentum recoil of $\pm\hbar K$ for the two pairs of light fields, with $ K \equiv (\omega_\mathrm{b}+\omega_\mathrm{r})/c$, and the atom gains kinetic energy $\hbar \omega_K$, in terms of the recoil frequency
	\begin{equation}
		\omega_K = \frac{\hbar K^2 }{2M},
	\end{equation} 
	where $M$ is the atomic mass. 
	
	The diffraction process is determined by the transferred energy, i.\,e. by the difference of the laser frequencies $\Delta \omega \equiv \omega_\mathrm{b} - \omega_\mathrm{r} $. A transition that is resonant for first-order diffraction corresponds in \cref{fig:FirstOrderSetup}(b) to the case where the solid arrows start and end on a parabola describing the kinetic energy of an internal state. This is possible if $\hbar \Delta \omega$ equals the kinetic energy  $\hbar \omega_K$ gained through recoil plus the energy difference $\hbar \omega_\mathrm{eg}$ between internal ground $\ket{g}$ and excited state $\ket{e}$, i.e.
	\begin{equation}
		\Delta \omega = \omega_\mathrm{eg}  + \omega_K.
		\label{eq:ResCondFirstOrder}
	\end{equation}
	Since the AC Stark shift can in principle be compensated, we refrain from including it in the subsequent discussion or the resonance condition. 
	
	The two pairs of light fields allow simultaneous diffraction in opposite directions but also enable spurious off-resonant transitions to higher diffraction orders denoted by dashed arrows. Additional couplings through polarization imperfections are neglected throughout this article.  Moreover, we assume plane waves and neglect wave-front distortions.
	
	The diffraction process depicted in \cref{fig:FirstOrderSetup}(b) is described by the truncated system of differential equations 
	\begin{subequations}
		\begin{align}
			\dot{g}_0 &= \, \I \Omega(t) \, \e^{-\I \omega_\mathrm{D} t} \, e_{1} 
			+ \, \I \Omega(t)  \, \e^{\I \omega_\mathrm{D} t} \, e_{-1}\\
			\dot{e}_{\pm 1} &= \, \I \Omega(t) \, \e^{\mp \I \omega_\mathrm{D} t} \, \e^{-\I 4\omega_\mathrm{K} t}  \, g_{\pm2} + \, \I \Omega(t) \, \e^{\pm \I \omega_\mathrm{D} t}  \, g_{0},
		\end{align}%
		\label{eq:FirstOrderEq}%
	\end{subequations}%
	coupling the ground state probability amplitudes $g_n \equiv g(p+n\hbar K)$ for the momentum eigenstate $\ket{p+n\hbar K}$ to the excited state amplitudes $e_n \equiv e(p+n\hbar K)$ with a (possibly time-dependent) coupling parameter $\Omega(t)$ that has a maximum amplitude $\Omega_0$.
	The system of equations is derived from the generalized version of the differential equations describing double Raman diffraction presented in Appendix~\ref{app:GeneralEquations}. Rabi oscillations take place between the probability amplitude $g_0 $ of the ground state and those of the excited state with two different momenta, $e_1 $ and $e_{-1} $. At the same time, the probability amplitudes of the excited states $e_{\pm1} $ couple off-resonantly to $g_{\pm2}$ indicated  with a detuning $4\omega_K$. These kind of transitions are prominent in the \textit{Raman-Nath} (\textit{Kapitza-Dirac}) regime \cite{muller2008atom,gould1986diffraction} where $\Omega_{0}/\omega_K \gtrsim 1$, but are suppressed in the \textit{Bragg}-type regime with $\Omega_{0}/\omega_K \ll 1$ in which double Raman is typically performed. Note that $e_{\pm2}$ couples further to higher diffraction orders, but these transitions are even more off-resonant and therefore suppressed. The Doppler frequency $\omega_\mathrm{D} = pK/M$ corresponds to the deviation from the resonant momentum $p_{0} =0$ within a wave packet and acts as a detuning to the resonant transition, leading to the effect of \emph{velocity selectivity} \cite{PhysRevA.45.342,PhysRevLett.66.2297,PhysRevLett.82.871,Szigeti_2012,PhysRevA.101.052512}. As coupling strength $\Omega(t) = \Omega_0 \, \mathrm{exp}[-t^2/(2\Delta \tau^2)]$ we consider a Gaussian function of width $\Delta \tau$.
	
	The coupling strength is connected to the pulse area $A$ via
	\begin{equation}
		A = \int \D t \,  \sqrt{2} \,\Omega(t).
	\end{equation}
	An area of $A= \pi/2$ leads to the transition $\ket{g,0} \rightarrow (\ket{e, \hbar K} + \ket{e, -\hbar K})/\sqrt{2}$, creating a superposition of left- and right-moving wave-packet components and therefore corresponds to a double-Raman beam splitter.
	
	\begin{figure}[tb]
		\includegraphics{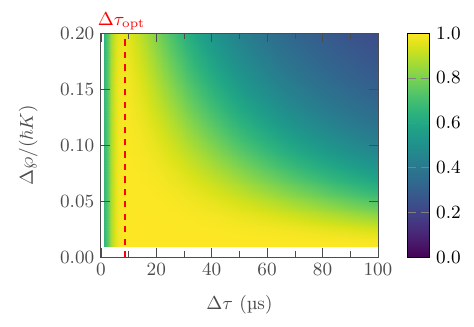}
		\caption{Efficiency $\mathcal{E}_{\pm 1}^{0}$ for a first-order double-Raman beam splitter for $^{87}$Rb as a function of the  width of the initial wave function $\Delta \wp$ and the pulse duration $\Delta \tau$ for a Gaussian pulse shape. Higher-order diffraction appears in the Raman-Nath regime (drop in efficiency for small $\Delta \tau$), for higher $\Delta \tau$ losses are determined by velocity selectivity. The dashed red line marks the optimal pulse duration.}
		\label{fig:BSfirstOrder}
	\end{figure}
	
	\subsubsection{Numerical treatment}
	\label{sec:NumericalTreatment}
	We numerically solve the system of differential equations \cref{eq:FullEqs} using the corresponding resonance condition \cref{eq:ResCondFirstOrder} for $^{87}$Rb with \textsc{Matlab}'s ODE45, a Runge-Kutta algorithm, with relative accuracy $10^{-3}$ and absolute accuracy $10^{-6}$. We calculate a transition function $G_{\Delta \tau}(p_\mathrm{f},p_\mathrm{i})$ which connects the initial and final momentum eigenstates \cite{PhysRevA.101.053610}. The transition function can be applied to the initial Gaussian wave packet $\psi_\mathrm{i}(p_\mathrm{i}) \propto \mathrm{exp}[-(p_\mathrm{i}-p_0)^2/(4\Delta \wp^2)]$ with $p_0=0$ to obtain the final wave function 
	\begin{equation}
		\psi_\mathrm{f}(p_\mathrm{f}) = \int \D p_\mathrm{i} \, G_{\Delta \tau}(p_\mathrm{f},p_\mathrm{i}) \, \psi_\mathrm{i}(p_\mathrm{i}). 
		\label{eq:psi_final}
	\end{equation}
	We truncate the range of momenta so that the solution for the diffraction efficiency (discussed in the following paragraph) obtained with $n_\mathrm{max}$ and $n_{\mathrm{max}+1}$ is at most of the same magnitude as the solver accuracy.

	\subsubsection{Diffraction Efficiency}
	We define the efficiency of an $n$th-order symmetric diffraction process between the momenta $\ket{\pm n_0 \hbar K}$ and $\ket{\pm ( n_0 + n) \hbar K}$ as
	\begin{equation}
		\mathcal{E}_{\pm n}^{n_0} = \int\limits_{p_{-}}^{p_{+}} \D p_\mathrm{f} \, \vert \psi_\mathrm{f}(p_\mathrm{f}) \vert^2 + \int\limits^{-p_{-}}_{-p_{+}} \D p_\mathrm{f} \, \vert \psi_\mathrm{f}(p_\mathrm{f}) \vert^2
		\label{eq:Efficiency}
	\end{equation}
	with the integration range $p_{\pm} = (n_0+n\pm 1/2)\hbar K$ and $n, n_0 \, \in \, \mathbb{N}$.
	Even though the expression works for arbitrary initial momenta, we have restricted ourselves to integer momenta $p_0 = n_0 \hbar K$ that are relevant for sequences of pulses.
	
	The efficiency $\mathcal{E}_{\pm1}^{0}$ of the first-order double Raman beam-splitter process sketched in \cref{fig:FirstOrderSetup}(b) is shown in \cref{fig:BSfirstOrder} as a function of the pulse duration $\Delta \tau$ and the width of the initial wave function $\Delta \wp$.
	For short pulse durations (i.\,e. in the Raman-Nath regime) diffraction into higher off-resonant orders becomes important and the efficiency of the beam-splitting process drops.
	For longer pulses, an efficiency close to unity demonstrates that diffraction in the Bragg-type regime leads to the targeted beam splitter. However, the longer the pulse, the more dominant the Doppler detuning becomes, which leads to velocity selectivity and the diffraction efficiency drops for broad momentum distributions. The red dashed line denotes the optimal pulse duration $\Delta \tau_\mathrm{opt}$ at intermediate times \cite{muller2008atom} and with highest efficiency for a broad range of different momentum widths $\Delta \wp$. For that, we determine for each value $\Delta \wp$ the pulse duration at which the maximal efficiency occurs and calculate the median over considered range of $\Delta \wp$, i.e. up to $0.2 \hbar K$. This time therefore describes the duration where the efficiency is good for various initial wave functions, however, for one realization with a particular momentum width a different value than $\Delta \tau_\mathrm{opt}$ can be better.
	It will later be used for a comparison between diffraction schemes.

	\subsection{Doppler-detuned Raman diffraction}
	\label{subseq:DopplerDetunedBox}
	\begin{figure}[tb]
		\includegraphics[width = \columnwidth]{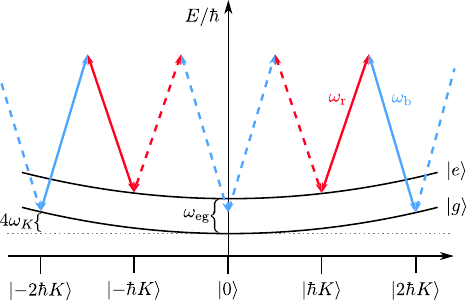}
		\caption{Energy-momentum diagram and resonant transitions for an atom with initial momentum $p_0 = \pm \hbar K$ (solid arrows) and initially in the excited state. The Doppler detuning of the spurious pair suppresses off-resonant transitions (dashed arrows), turning double into single diffraction. The initial conditions are chosen so that they correspond to a resonant sequential pulse following a double-diffraction beam splitter.}
		\label{fig:DopplerDetunedParabola}
	\end{figure}
	Atoms in a retroreflective setup  with initial momentum $p_0$ interact predominantly with only one of the two laser pairs because the other pair is Doppler detuned by $p_0 K/M$. Consequently, the double-diffraction process turns into a single-diffraction process, shown by solid arrows in \cref{fig:DopplerDetunedParabola}. Note that if the atom is in a superposition of momenta $\pm p_0$, two opposite but independent single-diffraction processes occur. However, the off-resonant Doppler-detuned transitions (dashed arrows from $\ket{e,\pm \hbar K}$ to $\ket{g,0}$ in \cref{fig:DopplerDetunedParabola}) are still present and cause a shift of the addressed atomic energy levels and by that detuned Rabi oscillations. A small detuning leads to the \emph{two-photon light shift} \cite{carraz_phase_2012,PhysRevA.94.063619}, while a large detuning reduces the diffraction efficiency. In contrast to Bragg diffraction, for Raman diffraction adiabatic elimination allows to identify the differential energy shift $\Delta E / \hbar$ for time-independent pulse shapes \cite{PhysRevA.78.043615}.
	For the momenta $p_0 = n_0 \hbar K$  with $n_0 \, \in \, \mathbb{N}$ that are of interest to our study of sequential pulses, one obtains for the widely used box-shaped pulses with Rabi frequency $\Omega_0$ the following differential energy shift
	\begin{equation}
		\Delta E / \hbar \equiv  \omega_K \delta= \pm \frac{\Omega_0^2}{\omega_K} \frac{2 n_0+1}{4n_0(n_0+1)} .
	\end{equation}
	The negative sign corresponds to transitions from $\ket{g,n_0 \hbar K}$ to $\ket{e,(n_0+1)\hbar K}$ while the positive sign corresponds to transitions from $\ket{e,n_0 \hbar K}$ to $\ket{g,(n_0+1)\hbar K}$. The detuning caused by this shift can be compensated by modifying accordingly the resonance condition from which $\Delta \omega$ is obtained. Box-shaped pulses are commonly employed for sequential Raman pulses, as they are easy to implement experimentally and have shorter durations compared to Gaussian pulses, while they can maintain a high diffraction efficiency.
	Note that the overall duration of a Gaussian pulse has to be truncated at a point which is significantly longer than its Gaussian temporal width $\Delta \tau$, so that the duration of a box-shaped pulse is small compared to the overall length of a Gaussian pulse, which we assume truncated to $8 \Delta\tau$.
	
	To demonstrate this effect, we consider in the following $p_0 = \pm \hbar K$, depicted in \cref{fig:DopplerDetunedParabola}, and $p_0 = \pm 2\hbar K$ for box-shaped pulses i.e. $\Omega(t)= \Omega_0$. The resonance condition for the transition $\ket{e, \pm \hbar K} \rightarrow \ket{g, \pm2 \hbar K }$, i.e. $p_0=\pm\hbar K$ as depicted in \cref{fig:DopplerDetunedParabola} is given by
	\begin{equation}
		\Delta \omega = \omega_\mathrm{eg}  - 3\omega_K + \Delta E / \hbar = \omega_\mathrm{eg}  - (3-\delta) \omega_K
		\label{eq:ResConBox1}
	\end{equation}
	with $\delta= 3 \Omega_0^2/(8 \omega_K^2)$.	The system of differential equations in an appropriate rotating frame reduces then to an effective two-level system without light shifts:
	\begin{equation}
		\left( \begin{array}{ccc}
			\dot{e}_{\pm 1}   \\                                              
			\dot{g}_{\pm 2}                                              
		\end{array}\right) =  \I \Omega_0 \left( \begin{array}{ccc}
			0 & \e^{\mp \I \omega_\mathrm{D}t}   \\                                              
			\e^{\pm \I \omega_\mathrm{D}t} &  0 \\                                            
		\end{array}\right) 
		\left( \begin{array}{ccc}
			e_{\pm 1}   \\                                              
			g_{\pm 2}                                               
		\end{array}\right)
		\label{eq:DopplerDetunedEq}
	\end{equation}
	Keeping in mind the differences between single and double diffraction, we now investigate $\pi$ pulses by choosing
	\begin{equation}
		A = \pi = 2 \,\Omega_0 \,\tau.
	\end{equation}
	Equation~\eqref{eq:DopplerDetunedEq} is analytically solvable, but to also calculate loss to off-resonant states that inevitably appears beyond the Bragg-type regime, we resort to a numerical treatment.
	
	Similarly, the resonance condition for the transition $\ket{g,  \pm2\hbar K} \rightarrow \ket{e,\pm 3\hbar K }$ with $p_0=\pm2\hbar K$ takes the form
	\begin{equation}
		\Delta \omega = \omega_\mathrm{eg} + (5+\delta) \omega_K
		\label{eq:ResConBox2}
	\end{equation}
	with  $\delta= - 5 \Omega_0^2/(24 \omega_K^2)$. It can be reduced to a two-level-system between $\ket{g, \pm 2\hbar K}$ and $\ket{e, \pm 3\hbar K }$ similar to \cref{eq:DopplerDetunedEq}.

	\begin{figure}[tb]
		\centering
		\includegraphics{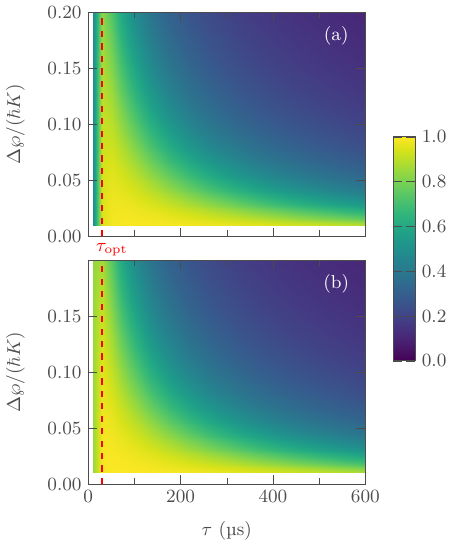}
		\caption{Efficiency for Doppler-detuned box-shaped Raman pulses  for varying width of the initial wave function $\Delta \wp$ and pulse duration $\tau$. In panel (a) we display the process $\ket{e,\pm \hbar K} \rightarrow \ket{g,\pm 2\hbar K}$ and in panel (b) the process $\ket{g,\pm 2\hbar K} \rightarrow \ket{e,\pm 3\hbar K}$. In the Raman-Nath regime both processes show transitions into other diffraction orders that reduce for increasing $p_0$, which makes the pulse in panel (b) more efficient. The dashed red lines mark the optimal pulse duration.}
		\label{fig:EfficiencyBox}
	\end{figure}
	Using the resonance condition Eq.~\eqref{eq:ResConBox1}, we calculate the diffraction efficiency $\mathcal{E}^{1}_{\pm 1}$  for the Doppler-detuned transition $\ket{e,\pm \hbar K}$ to $\ket{g, \pm 2\hbar K}$ and using the resonance condition Eq.~\eqref{eq:ResConBox2} to calculate the efficiency $\mathcal{E}^{2}_{\pm 1}$ for the transition $\ket{g, \pm 2\hbar K}$ and $\ket{e, \pm 3\hbar K }$ with an analogous numerical treatment as discussed in Sec.~\ref{sec:NumericalTreatment}. The only differences are the modified resonance conditions and box-shaped pulses i.e. $\Omega(t) = \Omega_0$. Moreover, the initial wave packet is a superposition of two Gaussians centered at $\pm p_0$ described by  
	\begin{equation}
		\psi_\mathrm{i}(p_\mathrm{i}) \propto \mathrm{exp}\biggl[-\frac{(p_\mathrm{i}- p_0)^2}{(4\Delta \wp^2)}\biggr]+ \mathrm{exp}\biggl[-\frac{(p_\mathrm{i}+p_0)^2}{(4\Delta \wp^2)}\biggr].
	\end{equation}

	Figure~\ref{fig:EfficiencyBox}(a) shows the efficiency for $p_0=\pm \hbar K$ and \cref{fig:EfficiencyBox}(b) the efficiency for $p_0=\pm 2\hbar K$ defined through \cref{eq:Efficiency} as a function of the width of the initial wave function $\Delta \wp$ and the pulse duration $\tau$.
	Although using different pulse shapes, we observe similar to \cref{fig:BSfirstOrder} diffraction to spurious orders in the Raman-Nath regime and therefore a significant loss of efficiency for short pulses. Since the spurious pair of light fields is increasingly off-resonant the larger the initial momentum~\cite{PhysRevA.101.053610}, the Raman-Nath regime is less important for the transition $\ket{g, \pm2 \hbar K} \rightarrow \ket{e,\pm 3\hbar K}$ compared to the transition $\ket{e, \pm \hbar K} \rightarrow \ket{g,\pm 2\hbar K}$.
	
	\begin{figure}[tb]
		\centering
		\includegraphics{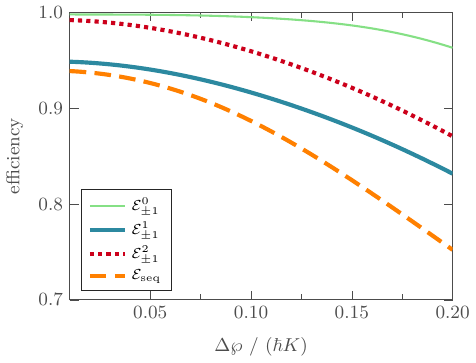}
		\caption{Diffraction efficiency for the three individual first-order pulses and their sequential application, obtained with the optimal pulse durations. The efficiency $\mathcal{E}_{\pm 1}^{0}$ for the double-Raman beam splitter with Gaussian pulse shape corresponds to the cut along the red dashed line in \cref{fig:BSfirstOrder}, the efficiencies of the Doppler-detuned and box-shaped single-diffraction mirror pulses $\mathcal{E}_{\pm 1}^{1}$ and $\mathcal{E}_{\pm 1}^{2}$ to the cuts along the red dashed lines in the two panels of \cref{fig:EfficiencyBox}. They differ because of the different diffraction geometries (Doppler-detuned single or double diffraction) as well as the pulse shape employed (Gaussian or box). The efficiency $\mathcal{E}_\mathrm{seq}$ of the sequential application of the three pulses is lower than the efficiency of the individual processes and obtained through the procedure explained in the text.}
		\label{fig:CutsFirstOrderSequence}
	\end{figure} 	
	We compare in \cref{fig:CutsFirstOrderSequence} the efficiency obtained with the optimal pulse duration $\tau_\mathrm{opt}\cong 30.7$\,\textmu s for the two effective single-diffraction pulses to that of the double-diffraction beam splitter (i.e., the cuts along the red dashed lines in Figs.~\ref{fig:BSfirstOrder} and~\ref{fig:EfficiencyBox}). Since the Raman-Nath regime is suppressed for Gaussian pulses, we observe that the double-diffraction beam splitter has the best efficiency for all momentum widths. Off-resonant couplings are suppressed by a Doppler detuning that scales with the initial momentum~\cite{PhysRevA.101.053610} and therefore affect the transition $\ket{e,\pm \hbar K}\rightarrow \ket{g,\pm 2\hbar K}$ more than the subsequent process with higher initial momentum. Hence, the first sequential pulse has the lowest efficiency of the individual pulses. However, these two diffraction types (single versus double diffraction) differ significantly in their geometry  as well as in the applied pulse shape, which makes a direct comparison difficult.

	\subsection{Three sequential Raman pulses}
	\label{subseq:Sequence}
	\begin{figure}[tb]
		\centering
		\includegraphics[width = \columnwidth]{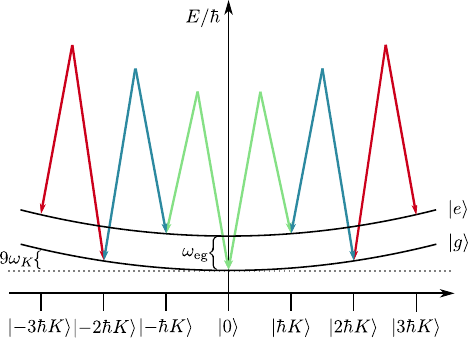}
		\caption{Energy-momentum diagram that shows the resonant processes of a sequence consisting of a double-diffraction beam splitter (green arrows) and two subsequent Doppler-detuned single-diffraction pulses (blue and red arrows). The initial wave packet is transferred from $\ket{g,0}$ to $\ket{e,\pm 3\hbar K}$ via the states $\ket{e,\pm \hbar K}$ and $\ket{g,\pm 2 \hbar K}$.}
		\label{fig:ComparisionSequence}
	\end{figure} 
	In this section we use the diffraction processes discussed in \cref{subsec:DoubleRaman,subseq:DopplerDetunedBox} to perform a Raman pulse sequence transferring population from the state $\ket{g,0}$ to an equal-amplitude superposition of $\ket{e,\pm 3 \hbar K}$. Raman pulses in a double geometry have already been experimentally realized, however only for the transition from $\ket{0}$ to $\ket{\pm 2\hbar K}$ \cite{PhysRevLett.103.080405}.
	A double-diffraction beam splitter with a Gaussian pulse shape transfers the initial wave function from $\ket{g,0}$ to $\ket{e,\pm \hbar K}$.
	Two subsequent box-shaped and Doppler-detuned effective single-diffraction Raman pulses transfer the population further to $\ket{g,\pm 2 \hbar K}$ and $\ket{e,\pm 3 \hbar K}$, see \cref{fig:ComparisionSequence}. The combination of Gaussian and box-shaped pulses in the sequence allows to benefit from their particular advantages regarding experimental duration and transfer efficiency. Each pulse induces first-order diffraction and requires an adjustment of the laser frequencies to fulfill the corresponding resonance conditions from \cref{eq:ResCondFirstOrder,eq:ResConBox1,eq:ResConBox2}. We use the optimal pulse durations $\Delta \tau_\text{opt} \cong 8.8\,$\textmu s and  $\tau_\mathrm{opt}\cong 30.7$\,\textmu s obtained in \cref{subsec:DoubleRaman} and \cref{subseq:DopplerDetunedBox} for the individual pulses.
	
	To calculate the overall efficiency, we use the initial wave function of width $\Delta \wp$ and apply a first-order double-Raman beam splitter.
	The diffracted wave function resulting from Eq.~\eqref{eq:psi_final} is used as initial condition for the next effective single-diffraction pulse.
	The obtained wave function, again calculated with the help of Eq.~\eqref{eq:psi_final} and the adjusted transfer function $G_{\tau_\text{opt}}$, is diffracted by the final pulse and the efficiency of the whole sequence is calculated through Eq.~\eqref{eq:Efficiency} by integrating over the population in the states $\ket{e,\pm 3\hbar K}$ determined by the final momentum distribution.
	This sequence of optimal pulses leads to a momentum transfer of $\pm 3\hbar K$ and its efficiency $\mathcal{E}_\mathrm{seq}$ is shown in Fig.~\ref{fig:CutsFirstOrderSequence}.
	Because the diffracted wave function after the first pulse has narrowed in momentum due to velocity selectivity so that diffraction with the next pulse is performed for an effectively colder sample, the efficiency is slightly larger than the product of the individual efficiencies $\mathcal{E}_\mathrm{\pm1}^0 \mathcal{E}_\mathrm{\pm1}^1  \mathcal{E}_\mathrm{\pm1}^2 $ calculated for perfect Gaussian initial wave functions. In any case, compared
	to the three individual pulses shown in the figure, the efficiency of the sequence $\mathcal{E}_\mathrm{seq}$ is lower. In fact, it is mainly limited by the lowest efficiency $\mathcal{E}_{\pm 1}^{1}$ of the first sequential pulse.

	\section{Third-order diffraction}
	\label{sec:ThirdOrderDiffraction}

	Instead of three sequential Raman pulses we focus in this section on only one pulse that relies on third-order diffraction to achieve the same momentum transfer of $\pm 3 \hbar K$.
	As \cref{fig:ParabolaThirdOrder} shows, the two laser pairs with frequencies $\omega_\mathrm{b}$ and $\omega_\mathrm{r}$ induce a six-photon diffraction process and transfer the population from $\ket{g,0}$ to $\ket{e,\pm 3\hbar K}$. The intermediate two-photon processes are off-resonant and thus, the states $\ket{e,\pm \hbar K}$ and $\ket{g,\pm 2\hbar K}$ are only virtually populated.
	
	\subsection{Resonance condition and pulse area}
	\label{subsec.ResCondandPulseArea}
	In such a third-order process, the atom gains due to its quadratic dispersion relation a kinetic energy of $ 9 \hbar\omega_K$, which leads for $p_0 = 0$ to the following modified resonance condition:
	\begin{equation}
		\Delta \omega =  \omega_\mathrm{eg} + (9+\delta) \omega_K
		\label{eq:ResCondThirdOrder}
	\end{equation}
	Here, we included the factor $\omega_K \delta$ to compensate for possible energy shifts similar to the Doppler-detuned diffraction processes in \cref{subseq:DopplerDetunedBox}.
	Since the pulses are time-dependent, we perform a numerical optimization of the efficiency and refer to Appendix~\ref{app.EnergyShifts} for a brief discussion of the analytical expression for box-shaped pulses.

	Because it is a third-order process and based on the results for box-shaped pulses, we expect the pulse area to scale with the third power of the Rabi frequency and the detuning with the second order.
	Hence, we obtain the energy shifts
	\begin{equation}
		\omega_K \delta =\beta \frac{ \Omega_0^2}{\omega_K}
	\end{equation}
	as well as the connection to the modified Rabi frequency and pulse area
	\begin{equation}
		A = \int \D t \, \alpha \, \frac{\Omega^3(t)}{\omega_K^2}
	\end{equation}
	through a numerical optimization of the diffraction efficiency with the \textsc{Matlab} function \texttt{fminsearch} by determining the optimization parameters $\beta$ and $\alpha$.
	For our range of initial momentum widths and pulse durations, we find that $\beta \, \in \, [-0.75,-0.42]$ and $\alpha \, \in \, [0.025,0.072]$ do not deviate much from the corresponding analytical value for box-shaped pulses given by \cref{eq:deltaBox,eq:RabiFreqBox}.
	
	\subsection{Comparison to first-order and sequential pulses}
	
	\begin{figure}[tb]
		\centering
		\includegraphics[width = \columnwidth]{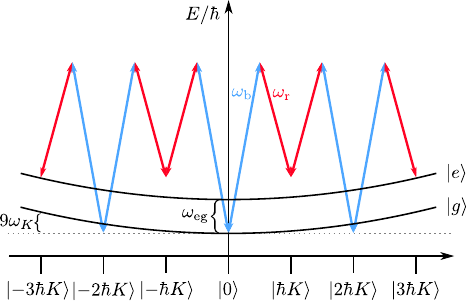}
		\caption{Energy-momentum diagram for a double Raman six-photon diffraction process. Two laser pairs (red an blue arrows) induce the transition. The first and second scattering process, each a two-photon process, are off-resonant. Thus, the transition from $\ket{g,0}$ to $\ket{e,\pm 3\hbar K}$ occurs by populating the states $\ket{e,\pm \hbar K}$ and $\ket{g,\pm 2\hbar K}$ only virtually.}
		\label{fig:ParabolaThirdOrder}
	\end{figure} 
	\begin{figure}[tb]
		\centering
		\includegraphics{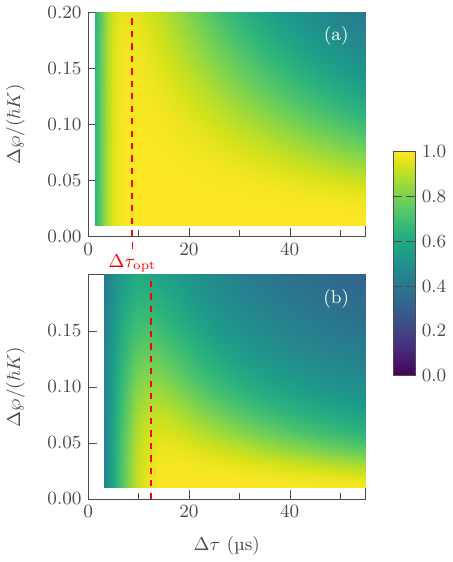}
		\caption{Comparison of the efficiency for a first-order double-Raman beam splitter $\mathcal{E}^0_{\pm1}$ (a)
			and a third-order beam splitter $\mathcal{E}^0_{\pm3}$ (a) for different widths of the initial wave function $\Delta \wp$ and pulse durations $\Delta \tau$.
			Panel (a) recalls the results for the first-order efficiency for times up to 60\,\textmu s from \cref{fig:BSfirstOrder} and panel (b) shows the simulated efficiency for third-order diffraction on the same time scale.
			While first-order diffraction is more efficient for a broad range of pulse durations in a Bragg-type regime, third-order diffraction is limited by two main effects: In the Bragg-type regime, higher-order diffraction is intrinsically limited by velocity selectivity, while for small pulse durations losses into intermediate state appear. The dashed red lines mark the optimal pulse duration.}
		\label{fig:EfficiencyOrder3}
	\end{figure} 
	
	We recall in \cref{fig:EfficiencyOrder3}(a) the efficiency of the first-order beam-splitter pulse $\mathcal{E}^0_{\pm1}$ from \cref{fig:BSfirstOrder} and compare it to the corresponding third-order beam splitter efficiency $\mathcal{E}^0_{\pm3}$ in \cref{fig:EfficiencyOrder3}(b) for different widths of the initial wave function $\Delta \wp$ and pulse durations $\Delta \tau$. As expected, third-order diffraction requires longer pulse durations, or, higher intensities for an efficient transfer since the population has to overcome two intermediate and off-resonant states. Moreover, velocity selectivity increases with the order of the diffraction process. Indeed, for $n$th-order diffraction the velocity spread associated with velocity-selectivity effects is proportional to $1/n$ because the effective Doppler detuning is given in that case by $n \omega_\mathrm{D} = n p K /m$. For small pulse durations losses into the intermediate states appear, especially into $\ket{g,\pm 2\hbar K}$ since it is the least off-resonant intermediate state as shown by \cref{fig:ParabolaThirdOrder}, while for larger pulse durations the loss of the efficiency of the diffracted population is mainly caused by velocity selectivity. Again, there exists a pulse duration $\Delta \tau_\text{opt}$ at which the atoms are diffracted most efficiently (red dashed line). When comparing these graphs, it seems that third-order diffraction is less efficient than the first-order pulse.

\begin{figure}[tb]
		\centering
		\includegraphics{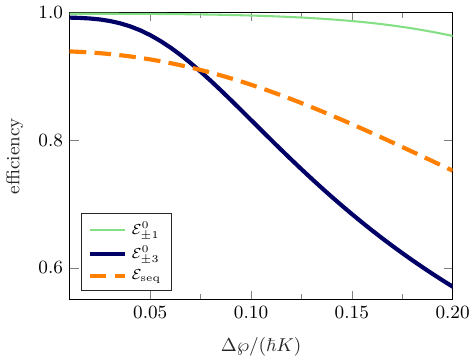}
		\caption{Efficiency at optimal pulse duration for different widths of the initial wave function $\Delta \wp$ for the first- and  third-order beam splitter as well as the sequence. Due to different regimes of pulse durations the first-order beam splitter is less velocity selective than its third-order counterpart. For small $\Delta \wp$ the third-order pulse is more efficient than the sequence, which makes it an interesting alternative for the diffraction of narrow wave packets like BECs. }
		\label{fig:CutsThirdOrderSequence}
	\end{figure}

	In \cref{fig:CutsThirdOrderSequence} we compare the efficiencies with optimal pulse duration for the first-order ($\Delta \tau_\mathrm{opt} \cong 8.8\,$\textmu s) and third-order beam splitter ($\Delta \tau_\mathrm{opt} \cong 13.3\,$\textmu s), as well as the Raman sequence introduced in \cref{subseq:Sequence} as a function of the widths of the initial wave function $\Delta \wp$. As already observed above, the first-order beam splitter has a higher efficiency than its third-order counterpart, which can be understood in terms of velocity selectivity and loss to intermediate states. However, if the targeted states are $\ket{e,\pm 3\hbar K}$, the third-order pulse has to be compared to the sequence of three first-order pulses rather than just the initial beam splitter.
	Indeed, the third-order pulse shows high efficiency for small momentum distributions, that exceeds the efficiency of the sequential application of three individual pulses. Even though the efficiency of the sequence could be improved by using Gaussian pulses throughout the sequence instead of only for the initial beam splitter, this would come at the cost of an even longer duration of the whole sequence.
	Consequently, third-order diffraction might be an interesting tool for the diffraction of wave packets with a narrow momentum distribution like BECs, since it allows to reduce the complexity of the experiment.
	In general, each transition of a sequence might introduce spurious phase contributions \cite{D_camps_2018} and using less pulses may facilitate the suppression of some uncertainties connected to frequency chirps \cite{PhysRevA.93.013609,PhysRevA.101.043622,PhysRevA.92.063617}. Furthermore, the overall duration of a single pulse can become shorter than that of a corresponding sequence of pulses, which might be particularly appealing for very compact setups \cite{Nelson2020} intended for real-world applications \cite{bongs2019taking}.

	\section{Mirror processes}
	\label{sec.Mirror_processes}
	
	So far we have focused on the discussion of third-order \emph{beam splitters} and saw that they can be an alternative to sequential Raman pulses.
	However, double-diffraction \emph{mirror} processes generally suffer higher losses to the intermediate momentum state around vanishing momentum compared to their single-diffraction counterpart so that their efficiency drops significantly for broad momentum distributions~\cite{PhysRevA.101.053610}.
	
\begin{figure}[tb]
		\includegraphics[width=\columnwidth]{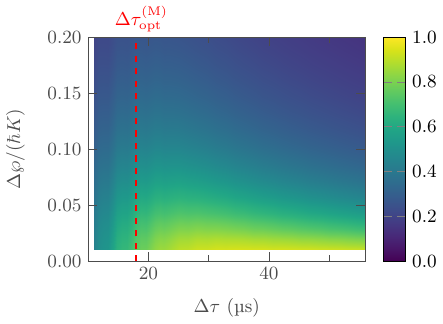}
		\caption{Efficiency $\mathcal{E}^{(\text{M})}$ for a third-order double-diffraction mirror as a function of the  width of the initial wave function $\Delta \wp$ and the pulse duration $\Delta \tau$ for a Gaussian pulse shape. The dashed line shows the optimal duration used for the following analysis.}
		\label{fig:MirrorDensity}
	\end{figure}

	To study the effect of a third-order double-diffraction mirror pulse, we center the initial wave packet around the momentum $p_0=-3\hbar K$ and define the efficiency through the population in the interval around $+3 \hbar K$, i.e.
	\begin{equation}
		\mathcal{E}^{(\text{M})} = \int\limits_{5\hbar K/2}^{7\hbar K/2} \D p_\mathrm{f} \, \vert \psi_\mathrm{f}(p_\mathrm{f}) \vert^2.
	\end{equation}
	Moreover, we adjust the pulse area to a mirror pulse by choosing $A = \pi$ and perform the optimization in analogy to Sec.~\ref{subsec.ResCondandPulseArea}.
	The resulting efficiency is shown in Fig.~\ref{fig:MirrorDensity} as a function of duration and initial momentum width.
	
\begin{figure}[tb]
		\includegraphics[width=\columnwidth]{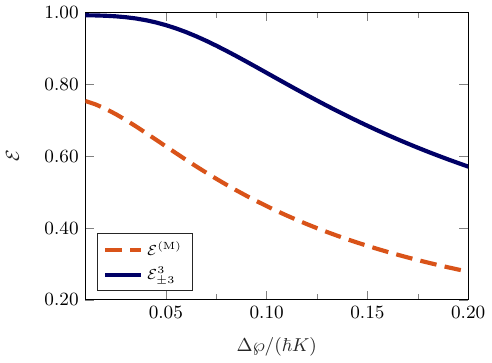}
		\caption{Efficiency at optimal pulse duration for different widths of the initial wave function $\Delta \wp$ for a third-order mirror (dashed orange) compared to a beam splitter (solid blue). The mirror process has a significantly lower efficiency than the beam splitter.}
		\label{fig:MirrorComparisonEffOptTau}
	\end{figure}

	While we observe again the effects of the Raman-Nath regime for a short duration and velocity selectivity for a long one, the details of the behavior are different from a beam splitter.
	This fact becomes more obvious when we again determine the optimal duration $\Delta \tau_\text{opt}^\text{(M)} = 17.97 $\,\textmu s (dashed line in the figure) and compare for different momentum widths the efficiency of beam splitter and mirror with their respective optimal durations, see Fig.~\ref{fig:MirrorComparisonEffOptTau}.
	Note that in our averaging procedure to obtain $\Delta \tau_\text{opt}^\text{(M)} $ large $\Delta \wp$ contribute equally so that they are overemphasized. As a consequence, for a particular width a different duration is preferable. Nevertheless, the
	mirror process is significantly less efficient over the whole range of investigated momenta and, similar to first-order diffraction, most of the population is lost to the state around vanishing momentum.
	Since these losses are similar to velocity selectivity, first-order double-diffraction mirrors are more efficient than their third-order implementation.
	
	\section{Interference signal}
	\label{sec.Interference_signal}
	
	\begin{figure}[tb]
		\includegraphics[width=\columnwidth]{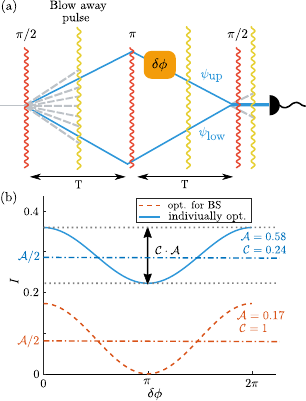}
		\caption{Space-time diagram of a third-order double-Raman Mach-Zehnder interferometer (a) and two corresponding interference signals (b). The resonant paths through the interferometer are represented by solid lines, the dashed off-resonant paths are removed, e.g., by blow-away pulses. The wave-packet component $\psi_\text{up}$ that has propagated along the upper path is phase-shifted by $\delta \phi$ with respect to the component $\psi_\text{low}$ that propagated along the lower path, before both interfere in the central exit port.
			The two exemplary interference signals in panel (b) are simulated with a width $\Delta \wp= 0.05 \hbar K$ and generated by individually optimized beam-splitter and mirror pulses (solid blue line) or by only optimizing the beam-splitter efficiency (orange dashed line), using the same duration and set of optimization parameters $(\alpha,\beta)$ also for the mirror pulse, solely doubling the pulse area. In the first case the amplitude $\mathcal{A}$ of the signal is larger, whereas the contrast $\mathcal{C}$ is well below unity, see the numerical values in the figure. In the second case, the preferable one, the amplitude drops, but the contrast remains unity.}
		\label{fig:interferometer}
	\end{figure}
	The third-order beam splitters and mirrors described so far have shown a higher velocity selectivity than their first-order counterparts, even with the optimization of the pulse parameters $(\alpha,\beta)$ discussed above.
	To investigate the potential and applicability of such processes for atom interferometry, we analyze as an example the idealized double-diffraction Mach-Zehnder interferometer shown in Fig.~\ref{fig:interferometer}(a), where the spurious (dashed) paths associated with off-resonant diffraction orders are disregarded, e.g. by the application of blow-away pulses.
	Although off-resonant paths that result from spurious first-order double Raman diffraction cannot be trivially eliminated by blow-away pulses, they do not contribute significantly to the interference signal and we neglect them in our approach. 
	We therefore study solely the effect of velocity selectivity and imprinted phases on the resonant diffraction orders, while we neglect the overlap with spurious paths in the exit port.
	The individual pulses are separated by a time $T$ during which both arms accumulate the phase $\mathrm{exp}\{-\mathrm i [p^2 / (2 m \hbar) + \omega_\mathrm{eg}] T\}$, where the first contribution describes the kinetic part and the latter accounts for the internal degree of freedom.
	For double Raman diffraction, the phase is the same for both arms and therefore cancels out.

	We simulate the interferometer in analogy to Ref.~\cite{PhysRevA.101.053610} and sequentially calculate the diffracted wave-packet components $\psi_\mathrm{up}(p)$ and $\psi_\mathrm{low}(p)$ evolving along the upper and lower path by using Eq.~\eqref{eq:psi_final} and the corresponding transition functions $G_{\Delta \tau}(p_\mathrm{f},p_\mathrm{i})$ for a beam splitter, a mirror, and a beam splitter.
	Introducing a phase shift $\delta \phi$, the interference signal in the central exit port can be expressed as
	\begin{equation}
		\begin{split}
			I(\delta\phi) &= \int \limits_{-\hbar K/2}^{\hbar K/2} \mathrm{d}p_\mathrm{f} \, \left| \psi_\mathrm{up}(p_\mathrm{f}) \, \mathrm{e}^{\mathrm{i} \delta\phi} +\psi_\mathrm{low}(p_\mathrm{f}) \right|^2 \\
			&= \frac{\mathcal{A}}{2} \left(1 + \mathcal{C} \cos \delta\phi\right)
		\end{split}
	\end{equation}
	with the amplitude $\mathcal{A}$ and the contrast $\mathcal{C}$.
	The amplitude reflects spurious diffraction and can be derived through $\mathcal{A} = \operatorname{max}[I(\delta \phi)] + \operatorname{min}[I(\delta \phi)]$, whereas the contrast $\mathcal{C} =( \operatorname{max}[I(\delta \phi)] - \operatorname{min}[I(\delta \phi)] )/ \mathcal{A}$ is determined by the asymmetry of the wave-packet components that evolved along the upper and lower path and asymmetrically imprinted phases.

	We show in Fig.~\ref{fig:interferometer}(b) an interference signal where we used the optimal duration of the mirror and the optimal duration of the beam-splitter pulses together with a momentum width of $\Delta \wp = 0.05 \hbar K$ (solid blue line).
	As expected from using the respective optimal durations, we observe a signal of large amplitude, although below unity.
	However, the contrast is imperfect, which we attribute to momentum-dependent phases imprinted asymmetrically across each wave-packet component.
	These phases occur because we use different optimization parameters $(\alpha,\beta)$ and durations for mirror and beam-splitter pulses.
	Indeed, when performing the same analysis by using $\Delta \tau_\text{opt}$ and the parameters $(\alpha,\beta)$ obtained from the optimization of the beam splitter also for the mirror, only doubling the pulse area, the interference signal has perfect contrast, see the dashed orange line in in Fig.~\ref{fig:interferometer}(b).
	Unfortunately, the amplitude drops significantly, since these parameters are not optimal for the mirror process, but otherwise no phases are imprinted asymmetrically.
	
	\begin{figure}[tb]
		\includegraphics[width=\columnwidth]{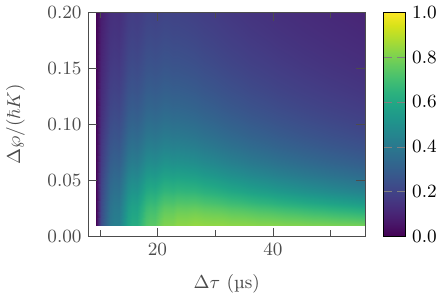}
		\caption{Amplitude $\mathcal{A}$ of the interference signal obtained from the joint optimization of beam splitters and mirror with one set of parameters $(\alpha,\beta)$ and the same pulse duration. We observe for each setting perfect contrast.}
		\label{fig:Amplitude}
	\end{figure}
	To overcome this issue, we do not perform an individual optimization of beam splitter and mirror, but optimize the amplitude of the interference signal using only one set of parameters $(\alpha,\beta)$ for both types of pulses that have the same duration and only differ in their pulse area by a factor of two.
	The resulting amplitude is shown in Fig.~\ref{fig:Amplitude} as a function of pulse duration and momentum width.
	Even though below unity, the amplitude reaches decent values for a broad range of parameters.
	Moreover, when we calculate the contrast $\mathcal{C}$, we see that it is unity for all durations and widths up to numerical precision.
	Hence, this simple interferometer simulation demonstrates that is it possible to preserve phase coherence and to observe a perfect interference pattern, where only the atom number in the exit ports of the interferometer is reduced due to velocity selectivity and diffraction into off-resonant orders.

	\section{Conclusions}	
	\label{sec:Conclusion}
	Double Raman diffraction allows in principle for resonant diffraction of odd orders, i.e. of order $2n+1$ with $n=0,1,2,\dots$\,. Higher diffraction orders come along with a higher velocity selectivity than first-order pulses. Moreover, higher intensities are necessary to overcome the intermediate states to achieve the optimal diffraction efficiency. However, when comparing third-order diffraction with a sequence consisting of three first-order pulses we find that third-order pulses diffract narrow momentum distributions like the ones associated with BECs more efficiently. The efficiency of our sequence, consisting of one Gaussian and two box-shaped pulses, could be improved by using Gaussian pulses only but at the cost of a significantly higher duration of the sequence.

	In contrast to double Bragg, double Raman diffraction allows a straightforward application of blow-away pulses, since the atom changes its internal state for a momentum transfer which is an odd multiple of $\hbar K$.
	Besides this drawback, we expect that a third-order double-Bragg beam splitter shows in principle a similar behavior of the diffraction efficiency. However, due to issues caused by the additional couplings inherent to double-Bragg diffraction, it is reasonable to assume that the overall efficiency will be lower than for its Raman counterpart.
	
	Third-order Raman mirrors can also be realized but suffer further limitations like losses into the central state $\ket{g,0}$, a feature intrinsic to double-diffraction mirrors~\cite{PhysRevA.101.053610}.
	Since double-Raman mirrors do not change the internal state similar to Bragg, one could in principle replace the mirror pulse through Bragg diffraction of sixth order from a standing wave to overcome such difficulties.

	However, using already different pulse parameters for a Raman mirror and beam splitters leads to a loss of contrast caused by momentum-dependent phases imprinted asymmetrically on both wave-packet components, so that the replacement of a Raman mirror by Bragg diffraction in an interferometer, besides the additional experimental complexity, is subtle and requires a careful analysis.
	Nevertheless, using the same pulse parameters for Raman mirror and beam splitters gives rise to a third-order double-Raman Mach-Zehnder interferometer with decent signal and good contrast.
	
	In addition to the possibility of higher-order diffraction, the symmetry of double-Raman pulses suppresses laser phase noise. Thus, it can be applied within LMT sequences together with Bragg diffraction or combined with Bloch oscillations.
	
	Hence, double Raman diffraction is a versatile tool for LMT techniques with the same flexibility and limitations as double Bragg diffraction. Not only does it occur naturally in microgravity or horizontal setups, it can also be combined perfectly with other LMT applications \cite{gebbe_twin-lattice_2019} to enhance the sensitivity of atom interferometers.
	
	\begin{acknowledgments}
		We thank M. Gebbe, M. Gersemann, C.~M. Carmesin, A. Friedrich and the whole QUANTUS group in Ulm as well as our partners of the QUANTUS collaboration for fruitful discussions, as well as J.-N. Siem{\ss} for helpful comments on our manuscript. This work is supported by the German Aerospace Center (Deutsches Zentrum f\"ur Luft- und Raumfahrt, DLR) with funds provided by the Federal Ministry for Economic Affairs and Energy (Bundesministerium f\"ur Wirtschaft und Energie, BMWi) due to an enactment of the German Bundestag under Grant Nos. DLR 50WM1556 (QUANTUS IV), DLR~50WM1956 (QUANTUS V), DLR 50WP1700 and 50WP1705 (BECCAL), 50RK1957 (QGYRO) as well as the Association of German Engineers ({Verein Deutscher Ingenieure}, VDI) with funds provided by the Federal Ministry of Education and Research ({Bundesministerium f\"ur Bildung und Forschung}, BMBF) under Grant No. VDI 13N14838 (TAIOL). E. G. thanks the German Research Foundation (Deutsche Forschungsgemeinschaft, DFG) for a Mercator Fellowship within CRC 1227 (DQ-mat). We thank the Ministry of Science, Research and Art Baden-W\"urttemberg (Ministerium f\"ur Wissenschaft, Forschung und Kunst Baden-W\"urttemberg) for financially supporting the work of IQ$^\mathrm{ST}$.
	\end{acknowledgments}

	\appendix
	\section{General equations}
	\label{app:GeneralEquations}
	In the following we discuss the differential equations for double Raman diffraction in their most general form, i.e. we do not focus on a specific resonance condition.
	A truncated version can also be found in Ref~\cite{ThesisLeveque}.
	
	The differential equations are derived within a rotating wave approximation \cite{schleich_quantum_2001} and the optically excited state is eliminated by adiabatic elimination \cite{bernhardt_coherent_1981,marte_multiphoton_1992,brion_adiabatic_2007}.
	Moreover, the equations are in an interaction picture with respect to the free evolution of the atoms and we assume that the laser phases vanish. They read
	\begin{subequations}
		\begin{align}
			\begin{split}
				\dot{g}_n = &\, \I \Omega(t) \, \e^{-\I [\omega_\mathrm{D} + \omega_\mathrm{eg} - \Delta \omega + \omega_\mathrm{AC} +(1+2n)\omega_\mathrm{K}]t} \, e_{n+1} \\
				+ &\, \I \Omega(t)  \, \e^{-\I [-\omega_\mathrm{D} + \omega_\mathrm{eg} - \Delta \omega + \omega_\mathrm{AC} + (1-2n)\omega_\mathrm{K}]t} \, e_{n-1}
			\end{split}\\
			\begin{split}
				\dot{e}_{n+1} = &\, \I \Omega(t) \, \e^{-\I [\omega_\mathrm{D} -\omega_\mathrm{eg} + \Delta \omega - \omega_\mathrm{AC} +(3+2n)\omega_\mathrm{K}]t} \, g_{n+2} \\
				+ &\, \I \Omega(t) \, \e^{-\I [-\omega_\mathrm{D} -\omega_\mathrm{eg} + \Delta \omega - \omega_\mathrm{AC} - (1+2n)\omega_\mathrm{K}]t} \, g_{n}.
			\end{split}
		\end{align}
		\label{eq:FullEqs}
	\end{subequations}
	Hence, the probability amplitudes of the ground state $g_n \equiv g(p+n\hbar K)$ and excited state $e_n \equiv e(p+n\hbar K)$ form a system of coupled differential equations. The coupling strength $\Omega (t)$ is determined by the laser intensity and the pulse shape. The frequency difference between ground state and excited state is given by $\omega_\mathrm{eg}$ while the Doppler frequency $\omega_\mathrm{D}$ denotes the deviation from a resonant momentum within a wave packet and thus, acts as a detuning. The recoil frequency is given by $\omega_K$ and the AC Stark effect by $\omega_\mathrm{AC}$. The adjustment of the laser frequency difference $\Delta \omega$ allows to perform resonant transitions between certain momentum states. Inserting the resonance condition \cref{eq:ResCondFirstOrder} into \cref{eq:FullEqs} leads for example to first-order diffraction as discussed in Sec.~\ref{sec:FirstOrderDiffraction}. A feature of double compared to single diffraction is the possibility to diffract into two directions simultaneously.

	\section{Energy shifts for box-shaped pulses}
	
	\label{app.EnergyShifts}
	
	Whereas we focus in the main body of our article on the compensation of energy shifts for Gaussian pulses, we discuss here briefly energy shifts that appear for third-order Bragg diffraction performed with box-shaped pulses.
	While energy shifts for higher-order \emph{single} Bragg diffraction can be derived through conventional adiabatic elimination of the intermediate states, the two counterpropagating pairs of light fields in double Raman diffraction prevent a straightforward application of the procedure \cite{paulisch2014beyond}, even though the technique can be generalized \cite{sanz2016beyond} to our case using Floquet theory.
	Similarly, we apply the method of averaging \cite{bogoliubov1961asymptotic,Kling_2018} that has already proven useful for double Bragg diffraction \cite{giese2013double} to provide the modified pulse area and energy shifts for box-shaped pulses.

	Inserting the resonance condition from \cref{eq:ResCondThirdOrder} into the system of differential equations given by \cref{eq:FullEqs}, leads with $\boldsymbol{v}\equiv(...,e_{n-1},g_n,e_{n+1},g_{n+2},...)$ and $\omega_\mathrm{AC}=0=\omega_\mathrm{D}$ to the coupled system
	\begin{equation}
		\dot{\boldsymbol{v}} = \I \Omega_0 \left(\mathcal{H}_0+ \sum\limits_{\nu \neq 0} \e^{\I \nu 2 \omega_\mathrm{K}t } \mathcal{H}_\nu \right) \boldsymbol{v},
	\end{equation}
	that, for time-independent $\Omega(t)=\Omega_0$, consists of the first-order contribution $\mathcal{H}_0$ and off-resonant couplings.
	The method of averaging constitutes an approach to systematically eliminate the time-dependent contributions $\mathcal{H}_\nu$ order by order of the adiabaticity parameter $\Omega_0/\omega_\text{K}$ to find an effective coupling between the desired states.
	Since we are interested in third-order processes, we have to perform this elimination up to third order.

	The internal states in $\boldsymbol{v}$ alternate and we find the dimensionless coupling matrices
	\begin{align}
		\begin{split}
			(\mathcal{H}_\nu)_{n,m}   = \e^{\I \delta \omega_\text{K} t}(\delta_{n+1,m} \delta_{4-n,\nu}+ \delta_{n-1,m} \delta_{4+n,\nu})\delta_{\ell,0} \\
			+  \e^{-\I \delta \omega_\text{K} t}( \delta_{n+1,m} \delta_{n+5,-\nu}+ \delta_{n-1,m} \delta_{n-5,\nu})\delta_{\ell,1}\\
		\end{split}
	\end{align}
	with the Kronecker symbol $\delta_{n,m}$ and $\ell $ the remainder of dividing $n$ by two.
	The first line describes the transition from the ground state to the excited state, while the second line the transition from the excited state to the ground state. 
	The first order of the method of averaging, i.e. $\Omega_0\mathcal{H}_0$, vanishes for the states of interest and the second given by $\Omega_0^2 \sum_{\nu \neq 0} \mathcal{H}_{-\nu}\mathcal{H}_{\nu}/(2\nu \omega_\mathrm{K})  $ vanishes as well for these transitions.
	However, the third order can be calculated through
	\begin{align}
		\begin{split}
			\label{eq.third-order-MoA}
			-&\frac{\Omega_0^3 }{12\omega_\mathrm{K}^2} \sum_{\substack{\nu,\sigma \neq 0 \\ \nu + \sigma \neq 0}} \frac{1}{\nu(\nu + \sigma)} [\mathcal{H}_{-\nu-\sigma},[\mathcal{H}_{\nu},\mathcal{H}_{\sigma}]] \\
			- & \frac{\Omega_0^3 }{8\omega_\mathrm{K}^2} \sum_{\mu \neq 0} \frac{1}{\mu^2} [\mathcal{H}_{\mu},[\mathcal{H}_{-\mu},\mathcal{H}_{0}]] 
		\end{split}
	\end{align}
	and contributes.
	With Eq.~\eqref{eq.third-order-MoA} it is possible to find an effective model that directly couples $g_0$ and $e_{\pm3}$ through
	\begin{equation}
		\begin{pmatrix}
			\dot e_{-3}\vphantom{\dfrac{o}{o}} \\
			\dot g_0\vphantom{\dfrac{o}{o}}\\
			\dot e_{+3}\vphantom{\dfrac{o}{o}}
		\end{pmatrix} = \I\omega_\text{K}
		\begin{pmatrix}
			\delta +\frac{5 \Omega_0^2}{16\omega_\mathrm{K}^2} & -\frac{\Omega_0^3}{32 \omega_\text{K}^3}  & 0\\
			-\frac{\Omega_0^3}{32 \omega_\text{K}^3} & -\frac{\Omega^2_0}{4\omega_\mathrm{K}^2} & -\frac{\Omega_0^3}{32 \omega_\text{K}^3}\\
			0 & -\frac{\Omega_0^3}{32 \omega_\text{K}^3} & \delta +\frac{5 \Omega_0^2}{16\omega_\mathrm{K}^2}
		\end{pmatrix}
		\begin{pmatrix}
			e_{-3}\vphantom{\dfrac{o}{o}}\\
			g_0\vphantom{\dfrac{o}{o}}\\
			e_{+3}\vphantom{\dfrac{o}{o}}
		\end{pmatrix},
	\end{equation}
	where we have transformed into an interaction picture by introducing a phase factor $\exp(\I \delta \omega_\text{K} t)$ to the elements $e_{\pm 3}$.
	We see an effective coupling $-\Omega_0^3/(32\omega_K^2)$ between different states that causes transitions and corresponds to the effective Rabi frequency.
	However, due to the asymmetry of the entries on the diagonal that effectively correspond to energy shifts, the Rabi-like oscillations are detuned.
	This detuning can be compensated by adjusting the resonance condition through
	\begin{equation}
		\delta = -\frac{9 \Omega_0^2}{16 \omega_K^2}.
		\label{eq:deltaBox}
	\end{equation}
	If these shifts are compensated and the oscillation is resonant, we find due to the three-level nature the pulse area
	\begin{equation}
		A  = \frac{\sqrt{2}\Omega_0^3}{32\omega_K^2} \tau.
		\label{eq:RabiFreqBox}
	\end{equation}
	
	\bibliography{bibfile}
	
	\clearpage
	
		\title[Erratum: Third-order double Raman diffraction]{Erratum: Atomic Raman scattering: Third-order diffraction in a double geometry [Phys. Rev. A 102, 063326 (2020)]\\[1ex]	
		\normalsize\normalfont{Published in \href{https://journals.aps.org/pra/pdf/10.1103/PhysRevA.106.029904}{Phys. Rev. A {\bfseries 106}  029904(E) (2022)}}} 
	
\makeatletter
\let\frontmatter@footnote@produce\relax
\makeatother
	
	\maketitle
	\thispagestyle{article}
	
		In the original paper we used an interaction picture with respect to the atomic evolution (consisting of center-of-mass motion and internal states) to calculate the transition function $G_{\scriptscriptstyle
		\Delta \tau}^{\text{\tiny(BS/M)}}(p_\mathrm{f},p_\mathrm{i})$ for a beam splitter (BS) or a mirror (M), which depends on the temporal width $\Delta \tau$ of the Gaussian pulse.
	The initiation of this interaction picture is chosen so that it coincides with the Schr\"odinger picture at the beginning of each pulse.
	
	When discussing an interferometer as a sequence of beam splitters and mirrors connected by the atomic free evolution in Sec.~V of our paper, the time evolution between the pulses was calculated in the Schr\"odinger picture. Moreover, the length of a Gaussian pulse was truncated to $\mathcal{T} = 8\Delta \tau$ and corresponded to the overall duration of the interaction picture.
	However, the atom-optical manipulations were not transformed back to this picture.
	This missing transformation leads to additional factors for each pulse.
	When working in the Schrödinger picture and after diffraction of the incoming wave function $\psi_{\mathrm i}(p_\mathrm{i})$ the final wave function $\psi_{\mathrm f}(p_\mathrm{f})$ in the relevant internal state $\ket{j}$ with $j \in \lbrace g, e \rbrace$, is given by
	\begin{equation}
		\psi_{\mathrm f}(p_\mathrm{f}) = \int_\mathcal{I} \mathrm  d p_\mathrm{i} 
		\e^{ -\I \bigl(\frac{p_\mathrm{f}^2}{2m\hbar} + \omega_j \bigl) \mathcal{T}}
		G_{\Delta \tau }^{\text{\tiny (BS/M)}}(p_\mathrm{f},p_\mathrm{i}) \, \psi_{\mathrm i} (p_\mathrm{i}).
	\end{equation}
	Here, $m$ denotes the atomic mass, and $\hbar \omega_j$ denotes the energy of the internal state after diffraction with $j \in \lbrace g, e \rbrace$.
	Because this inaccuracy led to a spurious momentum-dependent phase, the averaging process over all momenta that defined an exit port caused a loss of contrast that was unphysical.
	
	However, when all beam splitters and mirrors have the same duration, these spurious phase factors are common to both arms of the interferometer and cancel out, leading to the correct interference signal depicted by the orange line in Fig.~12(b) of the original paper.
	Similarly, the optimization of amplitude $\mathcal{A}$ in Fig.~13 of our paper corresponds to the physical interference signal, since the durations of all atom-optical elements are the same.
	
	In contrast, the transformation back to the Schr\"odinger picture does give rise to a non-trivial contribution if beam splitters and mirrors have different pulse durations.
	As a result, instead of the severe loss of contrast depicited by the blue line in Fig.~12(b) of the paper, the individually optimized pulse duration of the beam splitter and mirror leads to the interference signal displayed in Fig.~\ref{fig:interference_new}.
	This figure is the corrected version of Fig.~12(b) in the original paper.
	Even though it shows the same amplitude $\mathcal{A}$ as before, now for both situations the contrast is perfect.
	Indeed, using individually optimized pulse durations for the atom-optical manipulations leads to an increase in the signal's amplitude, but contrary to the original statement in our paper it does not influence the contrast.
	Thus, using $\Delta \tau_\mathrm{opt}^{\mathrm{(BS)}}$ for the beam splitter and $\Delta \tau_\mathrm{opt}^{\mathrm{(M)}}$ for the mirror is sufficient to implement an efficient third-order Raman interferometer with high contrast.
	The technique is therefore even better than expected from the results presented in our paper.
	
	\setcounter{figure}{0}
	\begin{figure}[htb!]
		\includegraphics[width=\columnwidth]{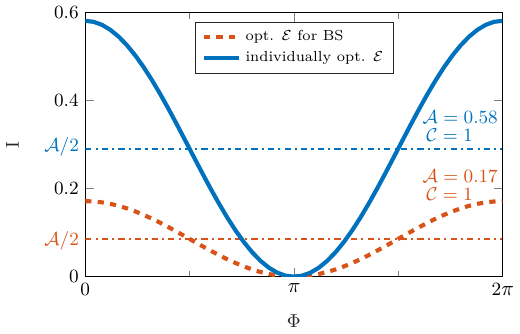}
		\caption{
			Two examples of interference signals obtained for a width $\Delta \wp = 0.05 \hbar K$ generated by the individually optimized beam-splitter and mirror pulses (solid blue line) or by only optimizing the beam-splitter efficiency (orange line) in complete analogy to Fig.~12(b) of our paper.
			In both cases we observe perfect contrast. However, using an optimized duration for both mirror and beam splitters leads to an significant increase in amplitude $\mathcal{A}$.
		}
		\label{fig:interference_new}
	\end{figure}
	
\end{document}